\documentclass[10pt,preprint2]{aastex}

\newcommand{\et}{{\it et~al.}}

\slugcomment{\apj \, submitted}
\shorttitle{Pulsar Wind Tori}
\shortauthors{Ng \& Romani}

\begin{document}

\title{Fitting Pulsar Wind Tori}

\author{C.-Y. Ng \& Roger W. Romani}
\affil{Department of Physics, Stanford University, Stanford, CA 94305}
\email{ncy@astro.stanford.edu, rwr@astro.stanford.edu}

\begin{abstract}
      {\it CXO} imaging has shown that equatorial tori, often with polar
jets, are very common in young pulsar wind nebulae (PWNe). These structures
are interesting both for what they reveal about the relativistic wind itself
and for the (nearly) model-independent measurement of the neutron star spin 
orientation they provide. The later is a particularly
valuable probe of pulsar emission models and of neutron star physics.
We describe here a procedure for fitting simple 3-D torus models to the 
X-ray data which provides robust estimates of the geometric parameters. 
An application
to 6 PWN tori gives orientations, PWN shock scales and post-shock wind speeds
along with statistical errors. We illustrate the use of these data by
commenting on the implications for kick physics and for high energy
beaming models.

\end{abstract}


\section{Introduction}

	The Crab nebula has long been known to have a subluminous zone and
termination shock surrounding its central pulsar. One of the striking successes
of the {\it Chandra X-ray Observatory} ({\it CXO}) mission has been to show that
this shock is likely an equatorial band and that similar structures are
seen around a number of young pulsars (e.g. Weisskopf \et\, 2000, Pavlov 
\et\, 2001, Helfand, Gotthelf \& Halpern 2001, Gotthelf 2001). 
Romani \& Ng (2003) argued that the apparent
symmetry of such PWNe, if interpreted as equatorial tori, allowed a 
robust fit for the 3-D orientation of the pulsar spin axis ${\vec \Omega}$ 
and showed that measurement of this axis can be effected even for quite 
faint PWNe. They also argued that, taking \ensuremath{\rm PSR\:J0538+2817}\
as an example, comparison of the spin axis with the proper motion axis
${\vec v}$ can be a sensitive probe of pulsar physics (see
Spruit \& Phinney 1998, Lai \et\, 2001). In particular, when the alignment
is due to rotational averaging of the transverse momentum, tight alignments
imply momentum kicks lasting many times the pulsar initial spin period. Such
time constraints on the kick timescale can exclude otherwise plausible models.
We wish here to systematize this comparison by outlining how robust fits 
for model parameters can be obtained even for relatively low-count 
{\it CXO} PWNe images.

	The characteristic scale of the wind termination shock around a pulsar
of spindown power ${\dot E}$ is
$$
r_{T} \approx ({\dot E}/4\pi c P_{ext})^{1/2}.
$$
This structure should be azimuthally symmetric about the pulsar spin axis when
$P_{ext} \ge P_{ram} = 6 \times 10^{-10}n v_7^2 {\rm g/cm/s^2}$, i.e. when the
pulsar is subsonic at speed $100v_7$km/s. Pulsars will seldom be sufficiently
slow to show toroidal shocks in the general ISM (where they will have PWN 
bow shocks), but young PSR often satisfy the azimuthally symmetric torus
condition in high pressure SNR interiors. \citet{vdS03} have provided 
analytic descriptions of conditions in SNR interiors, modeling the pulsar
parameters required for a subsonic PWN as the SNR evolves.

\section{Fitting Model and Technique}

The discovery papers showing obvious toroidal PWN structure have, of course,
often provided estimates of the torus scale, position angle and inclination.
However, most of these estimates were simple eyeball matches or, at best
2-D fits of ellipses in the image plane. The ubiquity of clearly
equatorial structures with toroidal symmetry and the robust physical
interpretation in terms of a static termination shock in an equatorial
wind argue that direct fits of the implied 3-D structure are most 
appropriate. Further, quantitative comparison with other pulsar observables
requires error estimates for the geometric parameters, which are often
not provided. Finally, we seek to constrain the orientations of central tori
in PWNe, even when they are not immediately obvious. All of these
considerations require a robust fitting procedure with (statistical)
error estimates. 

	We adopt a default model of a simple equatorial torus with a Gaussian
emissivity profile in cross-section. Clearly, this is not a sufficiently
detailed model to describe the complete PWN structure in the highest resolution
images. In fact, for Crab, Vela and a few fainter PWNe we find that multiple
tori component significantly improve the model fits. Also, we do not fit
the detailed radial variation of the torus emissivity and wind speed. Such
extensions of the model could be useful for comparison with numerical
models of relativistic shock flow (e.g. Komissarov \& Lyubarsky 2003,
Shibata {\it et al.} 2003). However, these
models are far from complete, and for the fainter PWNe the existing count
statistics do not warrant such model extensions. Accordingly we adopt the generic
torus fit and apply this to six pulsars, with increasingly poor image
statistics. Since for the bright PWNe fine structure beyond a simple torus
is visible, we must assume that these models are incomplete descriptions
of the fainter objects as well. However, the simple torus gives an adequate
description of the data and, assuming that this model captures the gross
nebular structure, we obtain quantitative estimates of the orientation
and wind flow parameters. The estimates are quite robust to perturbations
in the input fit assumptions. Note also that simpler structures are included
as subsets of our fit parameter space, e.g. a simple uniform halo, which
is reproduced by a face-on torus with small radius and large blur. Such
models are strongly statistically excluded, even for the faintest sources.

	We characterize the
termination shock (torus) by a polar axis at a position angle $\Psi$ 
($0-180^\circ$, measured N through E), with inclination angle 
$\zeta$ ($0-180^\circ$) from the observer line of sight.  The torus is 
simply described by a radius $r$ and a finite thickness or `blur' of 
Gaussian cross section, dispersion $\delta$, about the central ring. 
We assume that the surface brightness (averaged over its cross section)
is uniform in azimuth. However, the post-shock flow is still expected to be 
mildly relativistic and so an azimuthally symmetric ring will vary in 
apparent intensity, due to Doppler boosting. Synchrotron emission of intensity 
$I_0$ and photon spectral index $\Gamma$ in the rest frame will
have an apparent intensity
$$
I \propto (1-\mathbf{n}\cdot \beta)^{-(1+\Gamma)} I_0 
$$
\citep{pel87}
where $\beta=\mathbf{v}/c$ is the bulk velocity of the post shock flow
and $\mathbf{n}$ is the unit vector to the observer line of
sight.

	In practice we set up a coordinate grid with $x$, $y$
oriented with the CCD frame and $z$ along the observer line-of-sight.
This is rotated to a grid aligned with the torus
\begin{eqnarray*}
x^\prime&=&-x\cos \Psi -y\sin \Psi \\
y^\prime&=&(x\sin \Psi - y\cos \Psi)\cos \zeta + z\sin \zeta \\
z^\prime&=&-(x\sin \Psi - y\cos \Psi)\sin \zeta + z\cos \zeta
\end{eqnarray*}
with $x^\prime$ along the line of nodes, and $z^\prime$ along the
torus axis. In this frame the emissivity toward the Earth is
$$
I(x^\prime,y^\prime,z^\prime)= 
\frac{N}{(2\pi\delta)^2r} \times
$$
$$
\left(1 - \frac{y^\prime \sin \zeta}{\sqrt{{x^\prime}^2
+{y^\prime}^2}}\beta\right)^{-(1+\Gamma)}
e^{-\left[{z^\prime}^2+(\sqrt{{x^\prime}^2+{y^\prime}^2}-r)^2\right]/2\delta^2}
$$
where $N$ is normalized to the total number of counts in the torus
component.  We actually assume a fixed $\Gamma= 1.5$, typical of PWNe
and leave $\beta$ as a model parameter. 

	These parameters define the basic 3-D torus of PWN
emission. To form the 2-D torus image for data fitting, we
integrate the emissivity though the line of sight to get the 
counts in each pixel, i.e.
$$
C(x,y)=\sum_z I(x',y',z')
$$
The model image also generally
includes a uniform background and a point source for the pulsar,
whose intensity distribution is the PSF appropriate to the
detector position and source spectrum. For Crab and Vela,
the point sources have strong pile-up and we can optionally
blank a small region around the point source. There are a few
special cases of this basic model. In several PWNe (most notably
Vela), the termination shock has a double torus, presumably spaced
above and below the spin equator. We model this with identical tori,
symmetrically offset along the torus axis about the pulsar by 
distance $d$, following \citet{hgh01}.
In the case of the Crab, the structure includes two,
apparently inner and outer, tori. Finally in several cases
relatively bright polar jets complicate the analysis. We can
add model jet emission along the torus axis, blank the jet 
regions or include jet region photons as a fixed background.
We do not at present fit these components separately but 
their inclusion makes for more appealing model images. The torus parameters
are robust to the inclusion or exclusion of the `jet' region photons.
\begin{figure}[h!]
\scalebox{0.75}{\plotone{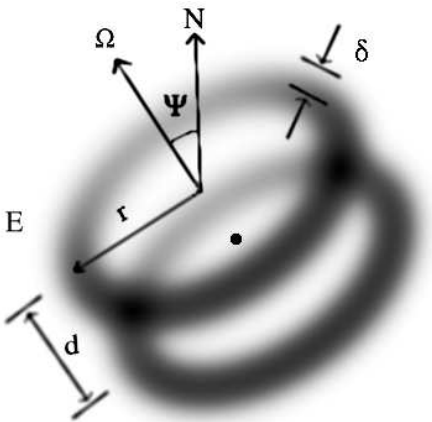}}
\caption{Torus Geometry -- The figure is the plane of the sky projection of
a double torus with fit parameters and angles labeled.  $\zeta$ (not shown)
is the angle between the sky plane normal (=the Earth line-of-sight $\hat z$) 
and ${\vec \Omega}$.\label{geom}}
\end{figure}

	Thus the basic torus model includes parameters $\Psi$, $\zeta$,
$r$, $\delta$, $\beta$ and a background. Additional parameters
can include the point source counts or double torus separation.
To keep the model parsimonious, we apply the constraint that the
total model counts $\Sigma_{xy} C_{xy}$ match the image counts 
(except for the Crab,
where bright complex background structure is not adequately
modeled by a constant background).  Our fitting procedure uses maximum 
likelihood, with a merit function formed from the summed log
probabilities of the observed counts $d_{xy}$ in each image
pixel. We use Poisson statistics 
$$
P(d_{xy}) = {{C_{xy}^{d_{xy}}~e^{-C_{xy}} } \over {(d_{xy}!)}},
$$ 
passing to Gaussian statistics
for an expected pixel count $C_{xy} \ge 20$ in the CCD image. In this
way the distribution of fluctuations in the fitted model parameters
passes smoothly to $\chi^2$  statistics in the limit of high count images.
The merit function is minimized, using the simulated annealing
package in Numerical Recipes \citep{pet92}. The initial parameter
estimates are chosen by eye and the simplex generally relaxes
quickly to the stable final solution, with a robust insensitivity
to the initial conditions. The one exception is the blur thickness
$\delta$, which has a tendency to grow to absorb the unmodeled
larger scale PWN enhancement present in many of the brighter nebulae.
As this parameter is not particularly useful in the physical
description of the shock, we fixed it at the initial (visual estimate)
value in a number of cases. The other fit parameters are quite
insensitive to $\delta$ over a broad range $\delta < r$.

Equally important are the parameter error estimates. We obtain these via
fits to Monte Carlo Poisson realizations of the best fit-model. After 500 random
simulations and fits we examine the distribution of the fit parameters.
These distributions are well-behaved, with near-Normal distributions
about the original fit parameters. We quote here 1-$\sigma$ confidence
intervals corresponding to 68\% CL. These are projected (i.e. true 
multi-parameter) estimates of the statistical errors in the fit parameters.
At very high confidence levels ($> 99$\%) the parameters become 
poorly constrained, especially for the fainter PWNe. Nevertheless
we find that the estimated statistical errors are quite reliable well
beyond 2-$\sigma$. Systematic errors, due to jets, complex backgrounds,
or other unmodeled structures are of course not included, but these
statistical estimates appear dominant for all but the brightest PWNe
and thus provide useful estimates of the parameters of the 
{\it torus component} which captures the gross structure of the central PWN.

The simulations also reveal correlations between the fitting parameters. 
$\Psi$ is, as expected, well decoupled from the other parameters.  $\zeta$ 
and $\beta$ are substantially co-variant with correlation coefficients 
as large as 0.7 since both affect the brightness ratio between the front 
and back sides of the tori. Fluctuations toward large inclination thus 
have reduced $\beta$. Similarly for the PWNe with large inclinations,
$\zeta$ and $r$ are covariant (coefficient 0.3 to 0.7) since the projected
angle of the torus' bright edge $r {\rm cos}\zeta$ is well constrained
by the images. Our reported 1$\sigma$ errors include these projected
correlations.

	Fit parameters and errors are reported for six PWNe with a range
of brightness in Table 1. We start with the `obvious' tori of Crab and Vela,
continue with the `apparent' tori of PSR J2229+6114 and PSR J1930+1852
and extend to two faint `plausible' tori from pulsars for which existing 
estimates of the proper motion make spin axis measurements of particular 
interest. There are perhaps 10 additional PWN known where such analysis
could be fruitful, and we expect there that several more useful torus 
measurements will be made as the {\it CXO} mission matures. In the present
analysis, all images are from archival ACIS-S
data sets, energy cut to 0.5-5keV. For the Crab we use ObsID 1998 (25ks),
for Vela we combine ObsIDs 1987 and 2813-2820 (180ks). For the other objects
we use the following observations: G54.1+0.3 (ObsID 1985, 40ks),
PSR J2229+6114 (ObsID 2787, 100ks), PSR B1706-44 (ObsID 757, 15ks),
PSR J0538+2817 (ObsID 2796, 20ks).  We mention here
special features of individual fits. 

	The two Crab tori are fitted sequentially, with the outer torus
fitted first and then used as a fixed background for the inner torus fits.
The model shown combines both components. The fitted tori are close to 
co-planar and have very small statistical parameter errors due to
the high image counts. The Vela tori are spaced symmetrically about the 
pulsar, centered along the symmetry axis; we quote the value $d$ in arcseconds
projected on the sky.  For the Crab and for Vela, the high S/N images 
show complicated fine detail in the tori and unmodeled fainter structures 
in the surrounding PWN. We have tested the fit sensitivity to these structures
by, eg. blanking the jet regions, adjusting the blur thickness and shifting
starting fit parameters. We find surprisingly small sensitivity to these changes
in the torus fits: the fit values are quite robust, although clearly
the very small statistical errors will be dwarfed even by the modest systematic
biases.

	For SNR G54.1+0.3 and PSR B1706-44, the global minima of the merit
functions pull the tori to small radii to absorb excess counts in the PSF wings
of the central pulsar. By excluding torus count contributions to the merit
function for pixels at $< 2^{\prime\prime}$, we obtain the minima
listed in the table, which match the visual structure of the nebulae.
PSR B1706-44 also has bright regions above and below the fit torus, which can
be interpreted as polar jets and which produce a
modest fraction of the counts attributable to the putative torus.
To minimize the sensitivity of the torus fit to these `jet' photons,
we add these counts as a fixed background in the model, as can be
seen in the corresponding model image. 

The PWN of PSR J0538+2817 is quite faint, providing only $\sim 2$\% of the 
point source counts. Accordingly, the geometry of the torus is sensitive to 
the amplitude of the PSF, and so the point source normalization cannot 
be fitted simultaneously. Instead the PSF amplitude is first fitted in
the central $1^{\prime\prime}$. This (pile-up corrected) PSF is held 
fixed during the torus fitting.
Consistency is checked by re-fitting for the point source in the 
presence of the torus background, confirming that the amplitude is 
unaffected. \citet{rn03} have previously described this PWN, arguing that the
spatial and spectral distinction from the PSF wings make the nebula highly
significant, despite its low count rate. The fit errors in Table 1 are slightly
larger that those in \citet{rn03}, because of the additional fit parameters
and the multi-dimensional errors. In particular, slightly smaller
errors ($\sim 6^\circ$ vs. $\sim 8^\circ$) can be obtained by direct fits 
to $\Psi$ alone. However, an equatorial extension of the PWN is strongly
required in the fit, the $\Psi$ estimate remains robust to perturbations
of the starting conditions and a simple uniform halo (small $\zeta$) is
strongly excluded -- the figure of merit degrades to 
the $3.6 \sigma$ level, referenced to our multi-parameter error flags,
for models with $\zeta \approx 0$.

\section{Application of Fit Parameters and Conclusions}

	In our fitting, we find that (for $\zeta$ not too small) the
position angle $\Psi$ is the most robust parameter. We believe that even if
the simple torus model is inadequate to fully describe all nebulae, this
measurement of the position angle of transverse PWN extension is very robust.
Thus comparison of the projected fit
axis with the proper motion axis remains an important application. For Crab, the
\citet{cm99} HST measurement of the Crab proper motion lies at
$\Psi_{PM}=292\pm10^\circ$.  This is $12^\circ$ or $1.2\sigma$ off of
our inner ring axis.  Similarly for Vela, we compare with the new radio 
interferometric proper motion of \citet{det03} at $\Psi_{PM}=302\pm4^\circ$, 
which has both higher astrometric accuracy and a corrected treatment of 
Galactic rotation effects from earlier optical estimates. This vector 
lies $8.6^\circ$ from our fitted torus axis, a $\sim 2.1\sigma$ discrepancy. 
Several other comparisons are semi-quantitative at present; we discuss
these below.

	Neither PSR J1930+1852 in G54.1+0.3 nor PSR J2229+6114 in G106.3+27
has a directly measured proper motion. Further, these SNR are complex with
no clear shell structure, so a velocity vector has not been estimated from
an offset birth site.  In contrast, \citet{dg02} and \citet{bg02} have 
re-examined the controversial association between PSR B1706-44 and G343.1-2.3,
and argued that this partly shell-like SNR is larger than previously believed,
placing the pulsar well in the interior and making the association more
plausible. The former authors find a southern extension suggesting a more 
complete circular shell; the vector from the center of this structure is 
at $\Psi_{PM} \sim 150^\circ$.  The latter authors suggest a proper motion 
parallel to the expansion of the nearby bright shell rim, $\Psi_{PM} \sim 
170^\circ$.  These position angles are $\sim 10-20^\circ$ from our fit axis, but
since the SNR evidently suffered asymmetric expansion, both of these
geometrical estimates are uncertain. Thus while the axes are
in general agreement, a direct measurement of the proper motion is essential 
for any serious comparison.  PSR J0538+2817, in contrast,
resides in S147 which has a quite symmetrical structure. \citet{rn03}
estimated a proper motion axis from the offset at $-32 \pm 4^\circ$, i.e.
$<1\sigma$ off of the PWN axis measured here. \citet{ket03} have recently
managed to extract a timing proper motion for this pulsar which confirms the
association with S147, although the PA is poorly determined. Here, both
higher statistics X-ray imaging and an astrometric proper motion
are needed to effect a precision test.

	There are two other young pulsars in shell SNR with
recent proper motion measurements, PSR B1951+32 in CTB80 
and PSR B0656+14 in the Monogem Ring.
Pulsar B1951+32 has a proper motion at $\Psi_{PM} = 252\pm 7^\circ$
away from its SNR birthsite \citep{met02} and is presently interacting
with the dense swept-up shell. Thus the prominent bow shock seen in HST
imaging is not unexpected; one would not expect a toroidal wind shock.
It is plausible that a `jet' wind could punch through the bow shock
and \citet{hest00} has proposed that the H$\alpha$ `lobes' bracketing the
bow shock mark the pulsar polar jets.  Under this interpretation we measure
the spin axis is at $\approx 265\pm5^\circ$, which is $13^\circ$ ($\sim 1.4
\sigma$) away from the proper motion axis.

With a new parallax distance measurement \citet{bri03} and \citet{thet03} find
that PSR B0656+14 is enclosed within the $\sim 66$pc-radius Monogem ring. The
surprisingly small proper motion at $\Psi_{PM} = 93.1\pm0.4^\circ$ implies
a transverse velocity of only 60km/s.  If the pulsar has a more typical $\sim
500$km/s space velocity, it must be directed along the Earth line-of-sight;
indeed, the parallax distance is consistent with the near side of the 
Monogem ring.
At its characteristic age, the pulsar should still be within the remnant interior
for radial velocities as large as $\sim 600$km/s. This SNR exploded in the low
density local ISM, so we expect the PWN to be toroidal with a characteristic
radius of $\sim 3^{\prime\prime}$. Interestingly, a short {\it CXO} observation
shows a faint, nearly circular halo around the pulsar at this scale \citep{pav02},
suggesting a face-on torus. Scheduled {\it CXO} observations have the sensitivity
to map this structure, which we predict will be consistent with a torus tilt
of $\sim 15^\circ$.

	Table 2 collects the projected proper motion and spin-axis 
position angle ($\vec v$, $\vec \Omega$) estimates along with estimates
of the line-of-sight inclination for these young pulsars. The trend
toward alignment (small $| \Delta \Psi_{\Omega \cdot v}|$) is strong, albeit 
imperfect. Formally, one should impose a prior on the maximum total
space velocity $v$ before evaluating the likelihood of a position angle 
range $\{\Delta \Psi\}$. If the 3-D angle between ${\vec v}$
and the projected position angle is $\theta$, then a physical upper limit
on the plausible $v=v_\perp {\rm cos \theta }/{\rm cos \Psi}$  restricts
the allowed range of $\theta$. However, in practice all of these pulsars
have relatively small $v=v_\perp$, so the disallowed range is negligible
and the probability of a position angle range is simply 
$2(\Delta\Psi_{max} -\Delta\Psi_{min})/\pi$.

If only the $1\sigma$ maximum $| \Delta\Psi |$ are considered, then for
Crab and Vela alone, there is a 3\% chance of obtaining alignments
as close as those seen from isotropically distributed 
${\vec v}, ~ {\vec \Omega}$. However if the other three angle estimates
of Table 2 are included, the chance probability falls to 0.04\%. On the
other hand, the weighted combination of these measurements gives
$\langle | \Delta \Psi_{\Omega \cdot v}| \rangle = 10.0 \pm 2.7 ^\circ$,
significantly different from 0. This finite misalignment
probes the characteristic timescale of the neutron star birth kick,
adopting the \citet{sp98} picture of rotational averaging. As discussed
in \citet{rn03}, the kick timescale constraints are very sensitive to
the initial spin periods of the individual pulsars. We defer a detailed
comparison with kick models to a later communication.

	A reliable measurement of the spin axis inclination
$\zeta$ can also be particularly valuable for these young pulsars.
Many of these objects are high energy (hard X-ray and $\gamma$-ray)
emitters and the modeling and interpretation of the pulse profiles
is quite sensitive to $\zeta$ \citep{ry95}. Radio techniques (polarization
sweeps and pulse width fitting) are often used to estimate $\zeta$,
but these are subject to substantial interpretation uncertainties.
For example polarization sweep results are affected by $90^\circ$ mode
jumps in various pulse components, which can often only be resolved
with single pulse studies. Perhaps this complexity is not surprising,
since the radio emission, at relatively low altitude, is sensitive
to higher order multipoles and the details of the magnetic polar cap
structure. In the wind zone all such details are likely lost. If
high energy emission is generated in outer magnetosphere ($\sim 0.1-0.3
r_{LC}$) gaps, then since $r_{LC}/r_{NS}$ is large, high order multipoles
should die away and the pulse profiles should be sensitive only to
magnetic inclination and $\zeta$.

	We list some radio inclination estimates in Table 2 -- however
we caution that substantially different values are available in the literature
for many of these pulsars. For Crab and PSR J0538+2817 the agreement with
our fit $\zeta$ appears good. For Vela and PSR B1706-44 the discrepancies are
large. Interestingly, \citet{hgh01} match the axis ratio of the projected
PWN torus by eye and find $\zeta = 55^\circ$ in good agreement with the
radio estimate; however this value is very strongly excluded in our fits. Our
relative $\zeta$s for Vela and PSR B1706-44 do however make sense in the
outer magnetosphere picture \citep{ry95}, with PSR B1706-44 at smaller $\zeta$
producing a narrower double $\gamma$-ray pulse and a larger phase delay
from the radio. We can further make some predictions for objects not
yet observed at $\gamma$-ray energies.
If our PWN $\zeta$ estimates hold up, we would expect $\gamma$-ray 
emission from PSR J2229+6114 to show a merged double pulse, somewhat
narrower than that of PSR B1706-44 (as appears to be the case in the 
hard X-ray), while PSR J0538+2817 should show a wide, Vela-like double pulse.
PSR J1930+1852 in G54.1+0.3 may be very faint in the $\gamma$-ray since its
small $180^\circ-\zeta=33^\circ$ suggests that any outer-magnetosphere
$\gamma$-ray beams miss the Earth line-of-sight. Finally PSR B0656+14 which
is many times fainter in $> 100$MeV $\gamma$-rays than expected from
its spin-down luminosity is widely believed to be viewed nearly pole-on.
In this orientation the strong outer-magnetosphere $\gamma$-ray beams
would not be visible, although we should see $\gamma$ emission from the 
pair production fronts in the radio zone above the polar cap. New
{\it CXO} imaging may allow a quantitative fit to the PWN, supporting the
apparent small $\zeta$.

	Also physically interesting are the estimates for the post-shock
velocities $\beta$. For the Crab nebula, our fit value compares well
with the $\beta \sim 0.5$ found for the motions of wisps near the torus
\citep{het02}, although clearly our very small statistical fit errors
must under-predict the true uncertainty. Moreover, it is puzzling that 
we get slightly larger $\beta$ for the outer ring. One certainly
expects the flow speeds to drop rapidly in the outer nebula and
{\it CXO/ HST}
data do show pattern velocities as small as $\beta \sim 0.03$
in the outer parts of the X-ray nebula. For Vela, no estimates of
$\beta$ from torus motions have yet been published, but \citet{pav03}
find $\beta \sim 0.3-0.6$ for features in the outer jet, which bracket
the torus $\beta$ found here. The interpretation of the outer jet speed is
apparently complicated by a varying orientation with respect to the line
of sight. Likely the inner jet/counter-jet provide a cleaner comparison
with the torus $\beta$; several authors have noted the larger counter-jet
brightness suggests that it is approaching the observer. Using the
time-averaged image in Figure 1, measuring a 5$^{\prime\prime}$ length
equidistant from the pulsar on each jet and subtracting the interpolated 
background from either side, we find a counter-jet/jet count ratio $f_B=2.3$.
For a continuous, time-average jet of photon index $\Gamma$ the
the Doppler boosting ratio is
$$
f_B = \left [ 
(1+\beta {\rm cos}\zeta)/(1-\beta{\rm cos}\zeta) \right ] ^{\Gamma+1}
$$
(the power $\Gamma+2$ applies for isolated bright blobs).  \cite{pav03}
report a inner jet spectral index $\Gamma\approx 1.1$, which with our fit
$\zeta$ gives $\beta_J = 0.45$, in remarkably good agreement with
our torus $\beta$. It is worth noting, however, that even these inner jets
show significant variation, so a time-resolved study of pattern speeds and
brightness is likely needed for a precise jet $\beta$. 

	No clear pattern yet emerges from the $\beta$ estimates presented here,
although there is a weak anti-correlation between the light cylinder magnetic
field and $\beta$ (correlation coefficient $\approx -0.3$). If significant,
this may suggest an anti-correlation between pair multiplicity, expected 
to be large in the narrow gaps of high ${\dot E}$, short period pulsars,
and the wind speed.  Ultimately, comparison of $\beta$ in different 
PWN components at different latitudes, and between PWNe, promises to
become an important new probe of pulsar electrodynamics. 


	For Vela, and a few fainter objects,
the presence of a double ring already suggests some increase in
pair multiplicity away from latitude $0^\circ$. If we de-project
the ring separation as seen from the pulsar, we get a brightness
peak (mid-plane of the torus) at co-latitude 
$$
\theta_{\rm Tor} = 
{\rm tan}^{-1} ( 2r{\rm sin}\zeta/d) \approx 74^\circ.
$$
This is about $10^\circ$ larger than our best fit $\zeta$. For Vela,
the polarization sweep rate maximum suggests a magnetic axis impact angle
$\beta=\zeta-\alpha = -4^\circ$.  Notice that this is on the same side
of the line-of-sight ($\alpha > \zeta$) as our fit torus. Thus we might
plausibly associate the two tori with a near-radial outflow of 
high pair multiplicity plasma from the magnetosphere open zones.
Perhaps further observation and modeling could distinguish between
pairs produced in a vacuum gap near the star surface (polar cap)
with plasma produced in an outer magnetosphere gap above the null-charge 
surface.
For example, the outer magnetosphere gaps should populate field
lines at angles $> \alpha$, consistent with the observed $\theta_{\rm Tor}> 
\alpha$.  If $\alpha$ is close to $\pi/2$ (e.g. Crab, PSR J0538+2817) the 
pair plasma from the two poles should merge, leading to a single thicker
torus.

\bigskip

	In conclusion, the ubiquitous azimuthally symmetric torus (+jet)
structures seen about young pulsars provide an important new probe of
the viewing geometry. We have described a procedure for fitting simple
geometric models to X-ray images that match the gross structure of the
central regions of many PWNe and provide robust estimates for model
parameters. These models clearly do not capture all of the rich, and dynamic,
structural details seen in the brighter nebulae, such as Crab and Vela.
However the fitting procedure does reduce biases of by-eye `fits' and does
allow extraction of geometrical parameters from quite faint objects.
Accordingly, such fits should allow us to measure spin axis orientations
for a larger set of pulsars and use these angles to probe models of
the neutron star kick and of pulsar magnetospheric physics.


\acknowledgments

This work was supported in part by CXO grant G02-3085X. We thank Walter Brisken
for discussion of recent pulsar astrometry.


\vfill\eject
\clearpage

\begin{deluxetable}{l|lllllll}
\tablehead{
object & $\psi$ & $\zeta$ & $r(^{\prime\prime})$ & $\delta$ & $\beta$ & PS/torus & sep($^{\prime\prime}$) \\ 
} 
\startdata
Crab (inner) & $124.0^{+0.1}_{-0.1}$ & $61.3^{+0.1}_{-0.1}$ & $15.60^{+0.03}_{-0.03}$ & 3.0* & $0.490^{+0.005}_{-0.006}$ & -/$1.0\times 10^5$ & -\\
\\
Crab (outer) & $126.31^{+0.03}_{-0.03}$ & $63.03^{+0.02}_{-0.03}$ & $41.33^{+0.02}_{-0.03}$ & 5.9* & $0.550^{+0.001}_{-0.001}$ & -/$1.1\times 10^7$ & -\\
\\
Vela & $130.63^{+0.05}_{-0.07}$ & $63.60^{+0.07}_{-0.05}$ & $21.25^{+0.03}_{-0.02}$ & 3.0* & $0.44^{+0.004}_{-0.003}$ & -/$1.3\times 10^6$ & $11.61^{+0.03}_{-0.03}$ \\
\\
SNR G54.1+0.3 & $91^{+4}_{-5}$ & $147^{+3}_{-3}$ & $4.6^{+0.1}_{-0.1}$ & $1.1^{+0.1}_{-0.1}$ & $0.62^{+0.04}_{-0.03}$ & 1701/602 & -\\
\\
PSR J2229+6114 & $103^{+2}_{-2}$ & $46^{+2}_{-2}$ & $9.3^{+0.2}_{-0.2}$ & 2.5* & $0.49^{+0.02}_{-0.02}$ & 2221/1113 & -\\
\\
PSR B1706-44 & $175^{+3}_{-4}$ & $55^{+3}_{-3}$ & $3.5^{+0.2}_{-0.1}$ & 0.74* & $0.65^{+0.03}_{-0.04}$ & 384/168 & -\\
\\
PSR J0538+2817 & $155^{+8}_{-8}$ & $99^{+8}_{-8}$ & $6.3^{+1.0}_{-0.7}$ & $1.7^{
+0.3}_{-0.7}$ & $0.54^{+0.09}_{-0.08}$ & 2442*/52 & -\\

\\
\enddata
\\
$^\ast$ = held fixed in the global fit.
\end{deluxetable}
\leftline{\qquad}

\begin{deluxetable}{llllll}
\tabletypesize{\normalsize}
\tablecaption{Proper Motion and Spin Axis Angles}
\tablehead{
$Pulsar$&$\Psi$ & $\Psi_{PM}^\dagger$ & $| \Delta \Psi_{\Omega \cdot v}|$ & $\zeta$ & $\zeta_{R}$ \\
}
\startdata
B0525+21& 	$124.0\pm0.1$ & $292\pm10$ & $12\pm 10$ & $61.3\pm0.1$ & 62$^a$  \\
B0656+14& 	-- & $93.1\pm0.4$ & -- & small & 16$^b$ \\
J0538+2817& 	$155\pm8$ & $328\pm\sim 4$ & $7\pm 9$ & $99\pm8$ & 97$^c$ \\
B0833$-$45& 	$130.6\pm0.1$ & $302\pm4$ & $8.6\pm 4$ & $63.6\pm0.1$ & 56$^d$ \\
B1706$-$44& 	$175\pm4$ & $160 \pm \sim 10$ & $15\pm \sim 11$ & $55\pm0.2$&75$^e$ \\
B1951+32& 	$85\pm\sim 5$ & $252 \pm 7$ & $13\pm \sim 9$ & -- & -- \\
\enddata
\\
References: $\dagger$ see text, $a$ see \citet{ry95}, 
$b$ \citet{lm88}, $c$ \citet{ket03}, 
$d$ \citet{kd83}, $e$ S. Johnston, priv. comm.
\end{deluxetable}

\begin{figure}[hb!]
\scalebox{2.2}{\plottwo{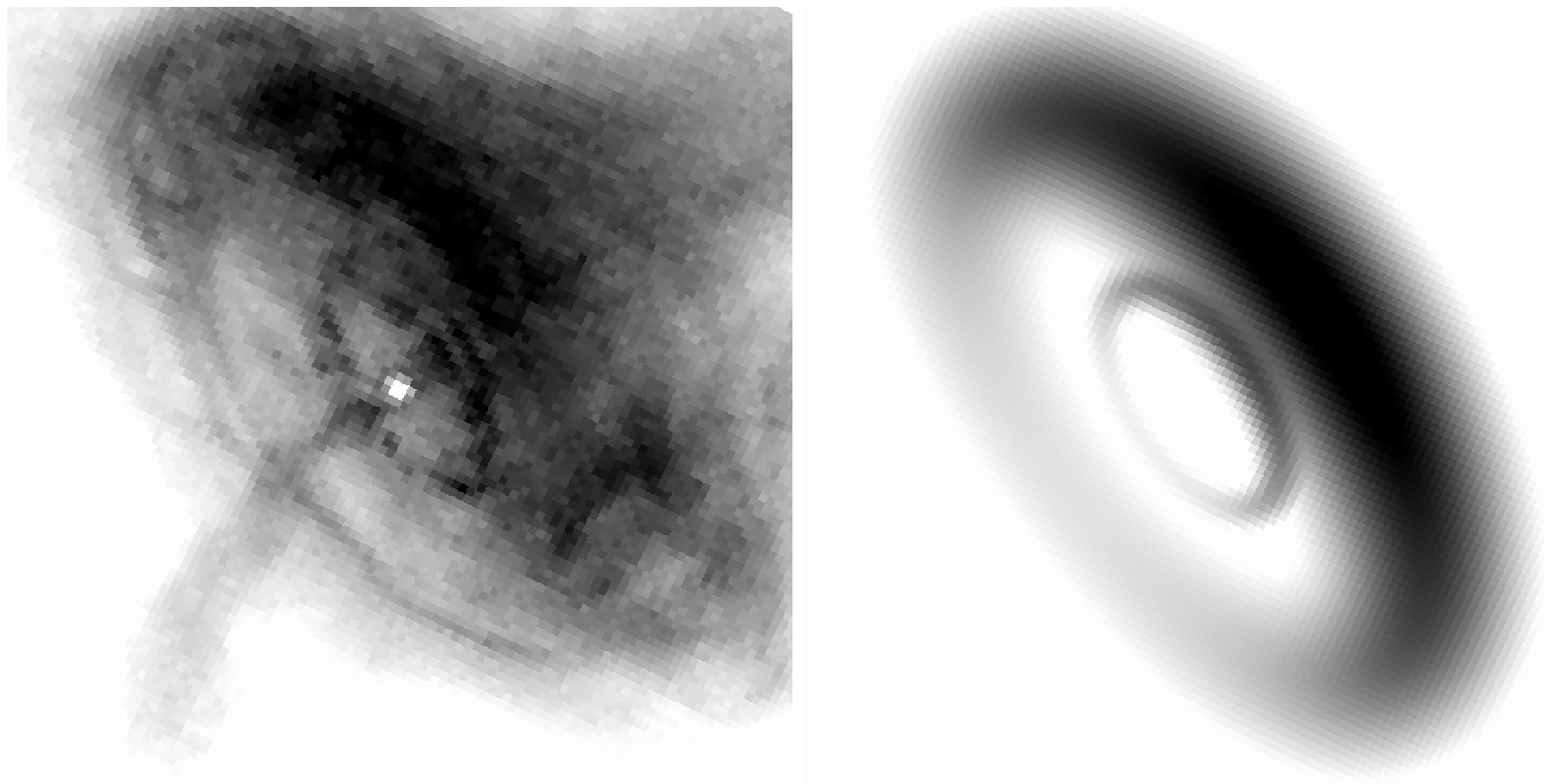}{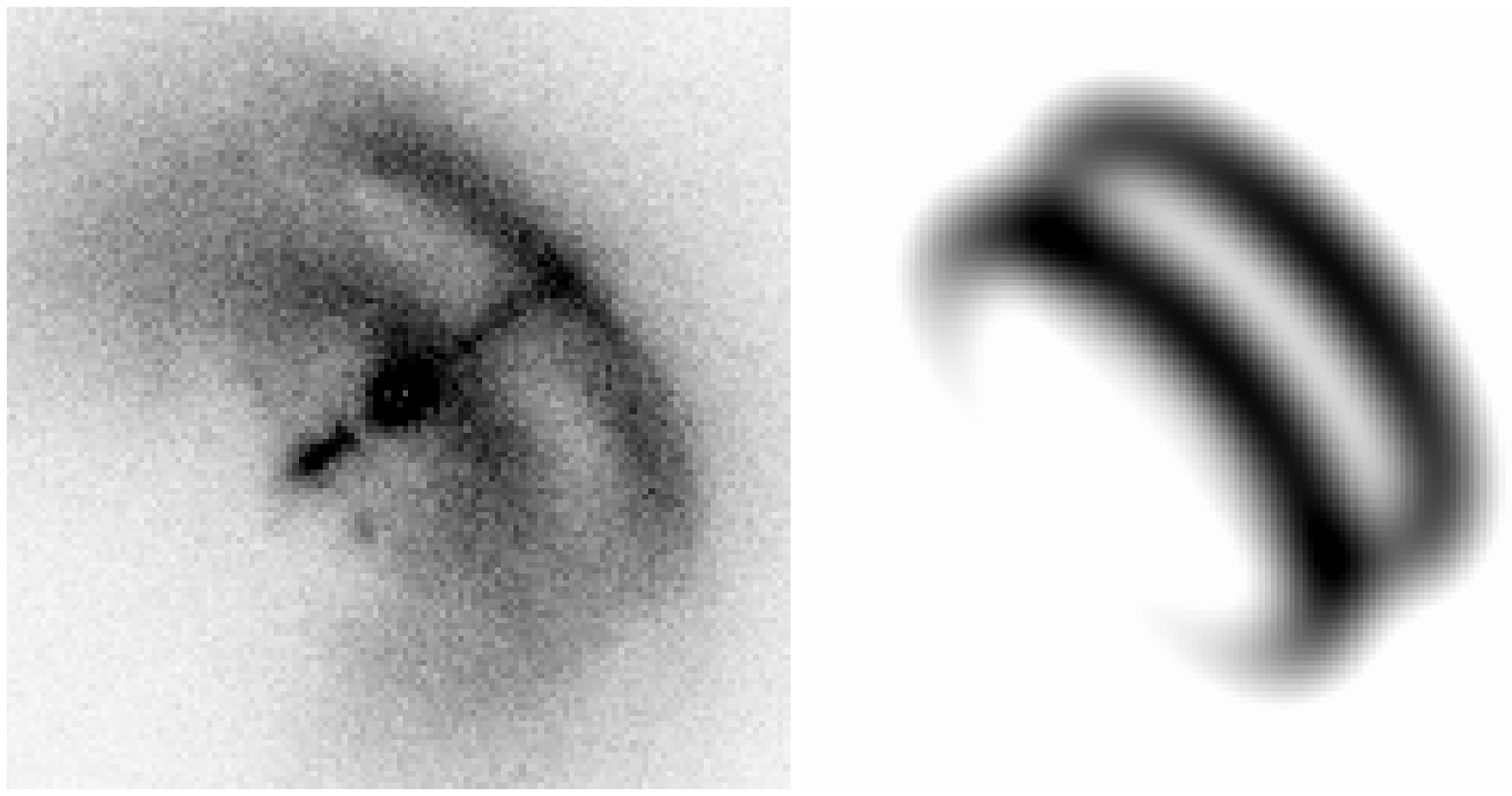}}
\scalebox{2.2}{\plottwo{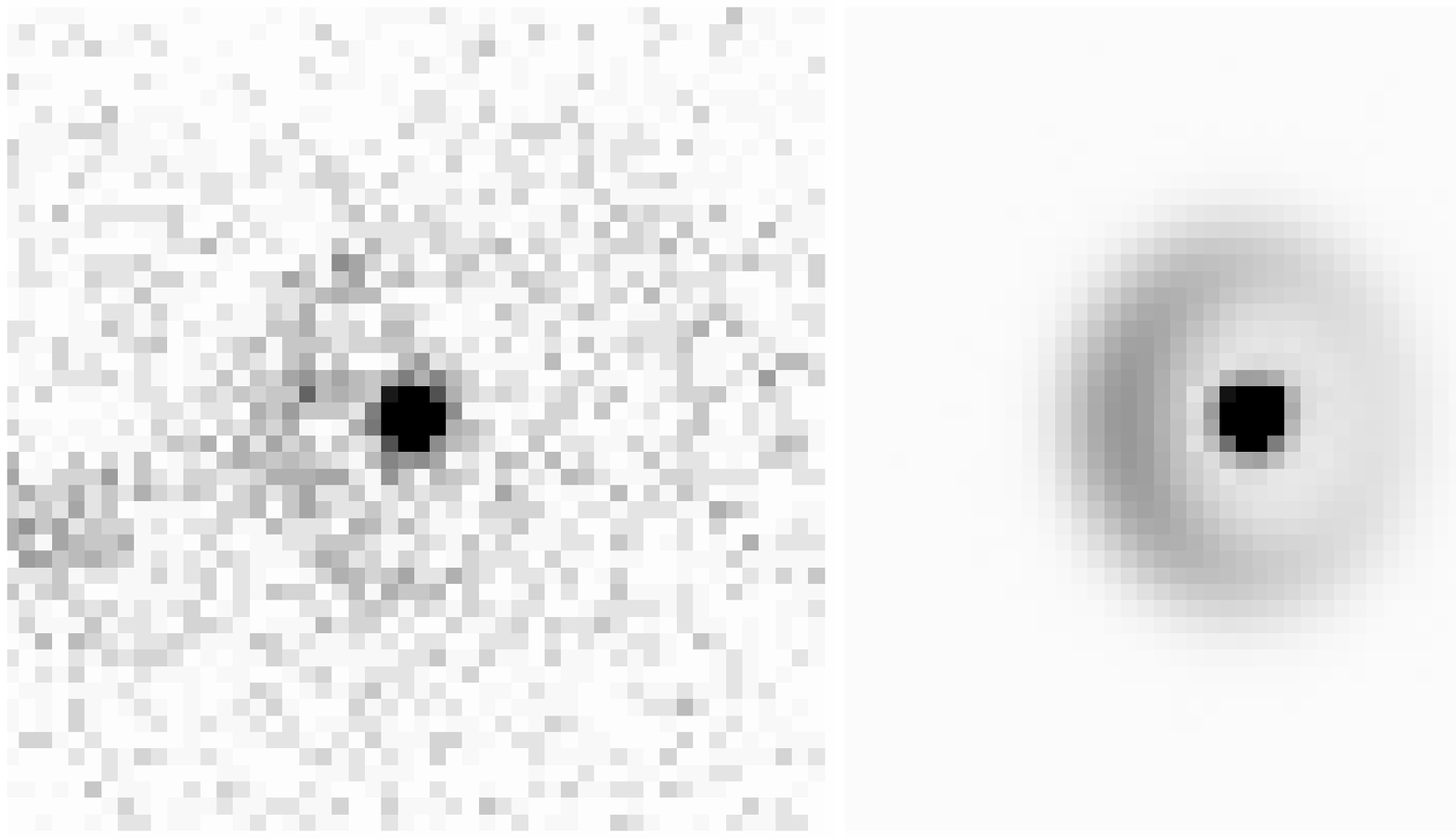}{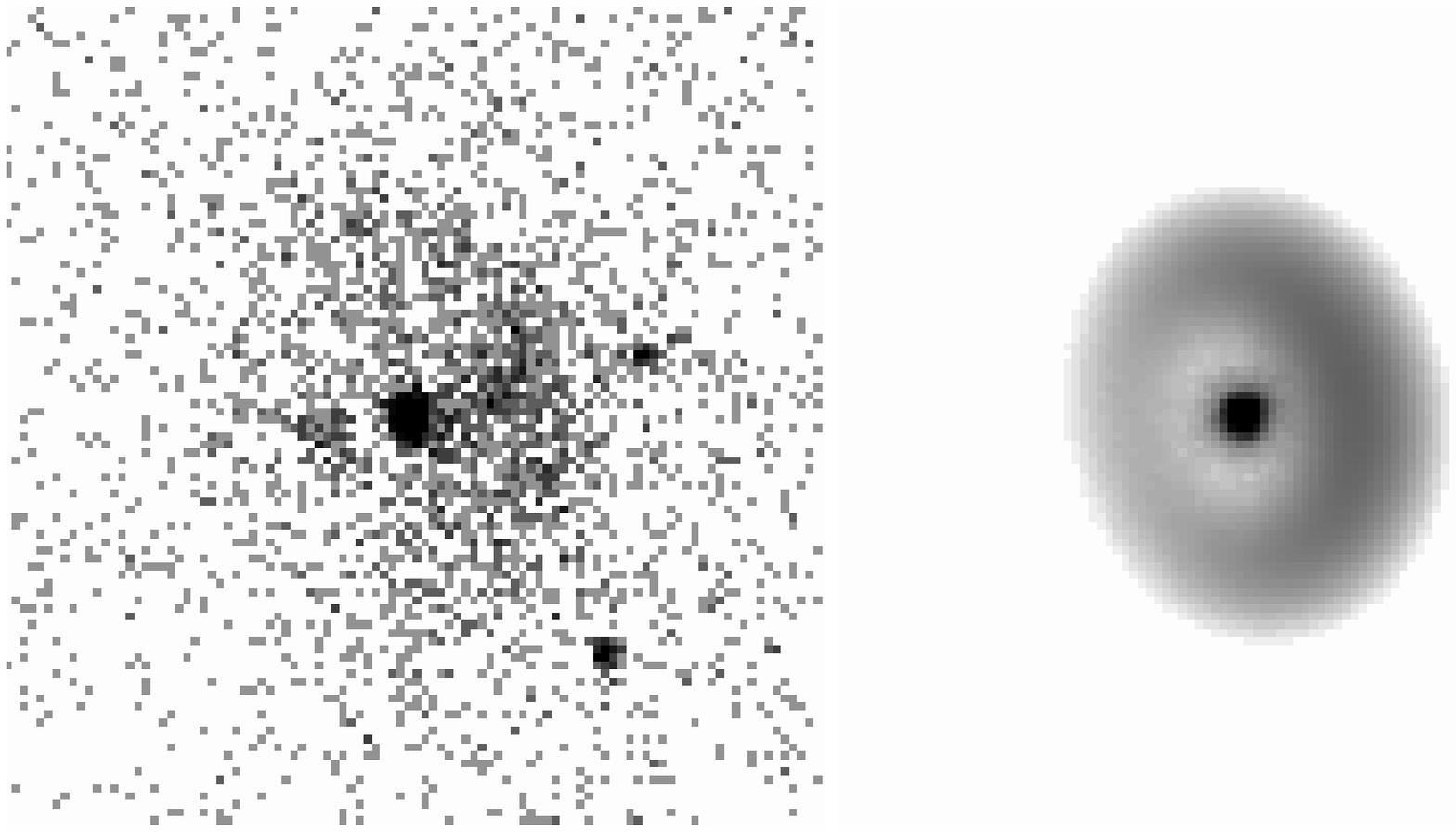}}
\scalebox{2.2}{\plottwo{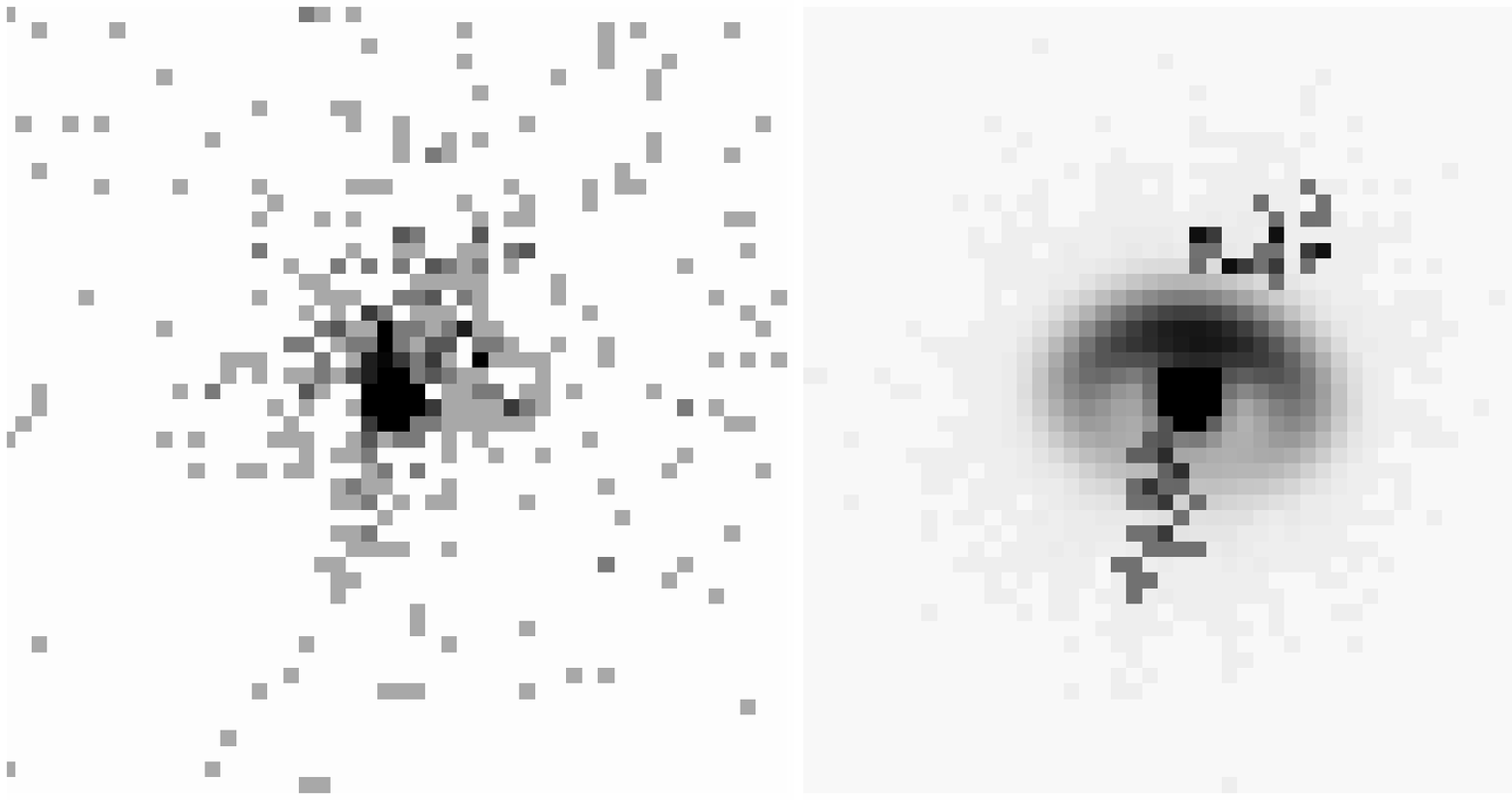}{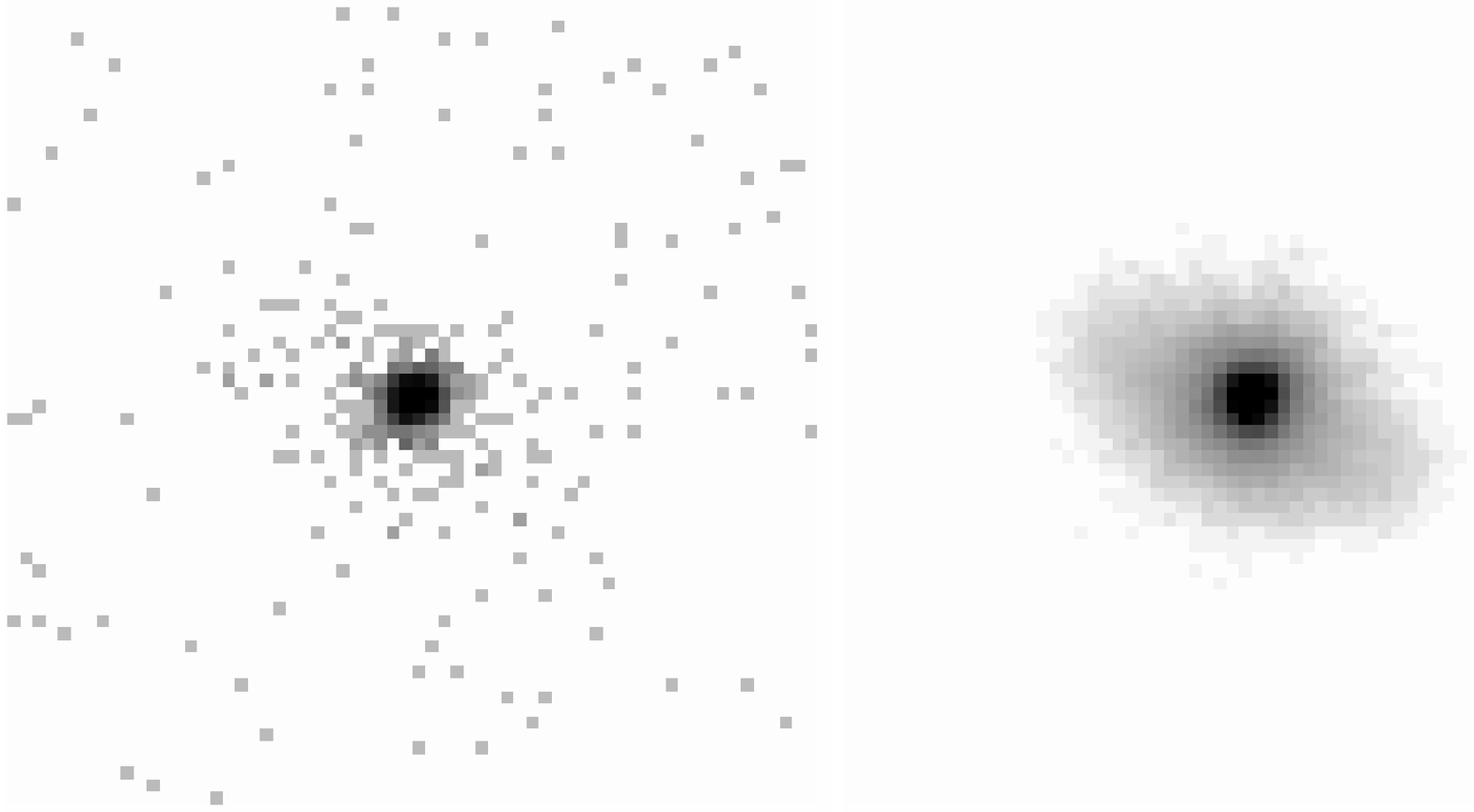}}
\caption{Pulsar Wind Tori -- {\it CXO} ACIS images (left) and best-fit models
(right) for Crab, Vela, G54.1+0.3, PSR J2229+6114, PSR B1706-44 and 
PSR J0538+2817 (left-right, top-bottom).\label{crab}}
\end{figure}


\begin{thebibliography}{}
\bibitem[Bock \& Gvaramadze (2002)]{bg02}Bock, D.C.-J. \& Gvaramadze, V.V. 2002,
\aap 394, 533
\bibitem[Brisken, \et (2003)]{bri03}Brisken, W., Thorsett, S.E., Golden, A. \& Goss, 
W.M. 2003, \aj, in press
\bibitem[Caraveo \& Mignani (1999)]{cm99}Caraveo, P.A. \& Mignani, R.P. 1999, \aap,
344, 357
\bibitem[Dodson \& Golap (2002)]{dg02}Dodson, R. \& Golap, K. 2002, \mnras, 334, L1
\bibitem[Dodson, \et (2003)]{det03}Dodson, R., Legge, D., Reynolds, J.E. \& McCulloch, P.M. 2003, astro-ph/0302374
\bibitem[Gotthelf (2001)]{g01}Gotthelf, E. V. 2001, AIP Conference Proceedings 
586, "Relativistic Astrophysics: 20th Texas Symposium." Ed. Wheeler and Martel
(Austin, Tx); astro-ph/0105128
\bibitem[Helfand, Gotthelf \& Halpern (2001)]{hgh01}Helfand, D.J., Gotthelf, E.V.
\& Halpern, J.P. 2001, \apj, 556, 380
\bibitem[Hester (2000)]{hest00}Hester, J.J. 2000, BAAS, 97.8216
\bibitem[Hester, \et (2002)]{het02}Hester, J.J. 2002, \apj, 557, L49
\bibitem[Komissarov \& Lyubarsky (2003)]{kl03}Komissarov, S.S. \& Lyubarsky, 
Y.F. 2003, \mnras, 344, L93
\bibitem[Kramer, \et (2003)]{ket03}Kramer, M., \et ~2003, \apj, 593, L31
\bibitem[Krishnamohan \& Downs (1983)]{kd83}Krishnomohan, S. \& Downs, G.S. 1983,
\apj, 265, 372
\bibitem[Lai, Chernoff \& Cordes (2001)]{lcc01}Lai, D., Chernoff, D.F. 
\& Cordes, J.M.  2001, \apj, 549, 1111
\bibitem[Lyne \& Manchester (1988)]{lm88}Lyne, A.G. \& Manchester, R.N. 1988, \mnras, 234, 477
\bibitem[Migliazzo \et (2002)]{met02}Migliazzo, J.M., \et ~2002, \apj, 567, L141
\bibitem[Pavlov, \et~ (2001)]{pksg01}Pavlov, G.G., Kargaltsev, O.Y., Sanwal, D. \&
Garmire, G.P. 2001, \apj, 554, L189
\bibitem[Pavlov, \et~ (2002)]{pav02}Pavlov, G.G., Zavlin, V.E. \& Sanwal, D. 
2002, in Proc. of 270th Heraeus Symp, 273.
\bibitem[Pavlov, \et~ (2003)]{pav03}Pavlov, G.G., Teter, M., Kargeltsev, O.
\& Sanwal , D. 2003, \apj, 591, 1157
\bibitem[Pelling, \et (1987)]{pel87}Pelling, R.M. \et ~1987, \apj, 319, 416
\bibitem[Press, \et (1992)]{pet92}Press, W.H., Flannery, B.P., Teukolsky, S. \&
Vetterling, W. 1992, {\it Numerical Recipes} (Cambridge University Press:Cambridge).
\bibitem[Romani \& Yadigaroglu (1995)]{ry95}Romani, R.W. \& Yadigaroglu, I.-A. 1995, \apj, 438, 314
\bibitem[Romani \& Ng (2003)]{rn03}Romani, R.W. \& Ng, C.-Y. 2003, \apj, 585 L41
\bibitem[Shibata, \et (2003)]{set03}Shibata, S., \et~ 2003, \mnras, in press.
\bibitem[Spruit \& Phinney (1998)]{sp98}Spruit, H. \& Phinney, E.S. 1998,
Nature, 393, 139
\bibitem[Thorsett, \et (2003)]{thet03}Thorsett, S.E., \et~ 2003, \apj, 592, L71
\bibitem[van der Swaluw {\it et al.} (2003)]{vdS03}van der Swaluw, E.,
Achterberg, A., Gallant, Y.A., \& Keppens, R. 2003, \aap, 397, 913 
\bibitem[Weisskopf, \et~ (2000)]{w00}Weisskopf, M.C. \et~ 2000, \apjl, 536, L81
\end{thebibliography}
\end{document}